\documentstyle{article}

\font\tenrm=cmr10
\font\tenit=cmti10
\font\elevenbf=cmbx10 scaled\magstep 1
\font\elevenrm=cmr10 scaled\magstep 1
\font\elevenit=cmti10 scaled\magstep 1

\font\ninerm=cmr9

\newcommand{\ftnotemark}[1]{${}^{#1}$}

\renewenvironment{thebibliography}[1]
 { \elevenrm
   \begin{list}{\arabic{enumi}.}
    {\usecounter{enumi} \setlength{\parsep}{0pt}
     \setlength{\itemsep}{3pt} \settowidth{\labelwidth}{#1.}
     \sloppy
    }}{\end{list}}

\begin{document}
\noindent MIT-CTP-2250  \hfill    hep-th/9311072\\
\noindent PUPT-1431 \hfill   October 1993\\
\begin{center}
{{\elevenbf
TWO-DIMENSIONAL QCD AND STRINGS\footnote{
\ninerm\baselineskip=11pt Based on a talk given by W. T.
at the Strings '93 conference, Berkeley, May 1993.  This work was
supported in part by the divisions of Applied Mathematics of the U.S.
Department of Energy under contracts DE-FG02-88ER25065 and
DE-FG02-88ER25066 and in part by the National Science Foundation under
grant PHY90-21984.  }\\}}
\vglue 0.58cm
{\tenrm  DAVID J. GROSS \\}
\baselineskip=13pt
{\tenit Joseph Henry Laboratories,  Princeton University\\}
\baselineskip=12pt
{\tenit Princeton, New Jersey  08544, USA\\}
\vglue 0.3cm
{\tenrm and\\}
\vglue 0.3cm
{\tenrm  WASHINGTON TAYLOR\\}
\baselineskip=13pt
{\tenit  Center for Theoretical Physics,  Laboratory for Nuclear
Science and Department of Physics\\}
\baselineskip=12pt
{\tenit  Massachusetts Institute of
Technology; Cambridge, Massachusetts 02139, USA\\}
\vglue 0.6cm
{\tenrm ABSTRACT}
\end{center}
\vglue 0.1cm
{\rightskip=3pc \leftskip=3pc \tenrm\baselineskip=12pt \noindent A
review is given of recent research on two-dimensional gauge theories,
with particular emphasis on the equivalence between these theories and
certain string theories with a two-dimensional target space.  Some
related open problems are discussed.}
\vglue 0.58cm
{\elevenbf\noindent 1. Introduction}
\vglue 0.38cm
\baselineskip=14pt
\elevenrm
There has been a recent renewal of interest in two-dimensional gauge
theories.  In two dimensions any pure gauge theory is locally trivial
and has no propagating modes.  However, by either considering the
theory on a compact 2-manifold or introducing external Wilson loop
sources, a nontrivial character of the theory emerges which is almost
topological in nature.  These two-dimensional theories have
provided an interesting simplified model with which to study certain
properties of gauge theories in any dimension.

One aspect of gauge theories which has long been of interest is the
connection between a gauge theory in $d$ dimensions and a string
theory with $d$-dimensional target space.  Speculations about how such
a connection might be made have motivated a wide range of research in
both QCD and strings.  In this talk I will describe some recent work
in which this connection is made rigorously for two-dimensional gauge
theories.  The string theories which will be discussed are of a very
special type, being described by covering maps {}from a string
world-sheet onto the two-dimensional target space with a finite number
of singularities and no folds.

I will begin by giving a brief review of some of the previous research
on two-dimensional gauge theories which is relevant to the work at
hand.  I will then give a simplified description of the proof that
these gauge theories are equivalent to string theories.  Following
this, I will discuss a variety of recent related work, and then close
with a brief description of several outstanding problems in this area.
\newpage
{\elevenbf\noindent 2. History}
\vglue 0.38cm
Almost 20 years ago\ftnotemark{1}, 't Hooft pointed out that in
certain circumstances it is reasonable and interesting to expand the
partition function and correlation functions of a $U(N)$ gauge theory
in powers of $N$.  The guiding principle behind this expansion is that
when one represents the gluon propagator by double lines corresponding
to the adjoint representation of the gauge group, each Feynman diagram
has associated with it a natural genus $g$; it turns out that the
power of $N$ associated with a Feynman diagram of genus $g$ is
precisely $N^{2-2g}$.  As a particularly simple example, 't Hooft
considered the case of 2-dimensional QCD.  He included fermionic
matter fields in the fundamental representation of the gauge group
(quarks), and was able to calculate the masses of mesons in the theory
to leading order in $N$ (the planar approximation).

In the following years, further progress was made in understanding the
large $N$ meson spectrum, and in connecting the two-dimensional QCD
theory to a stringy theory.  Bars and Hanson\ftnotemark{2}
showed that in the large $N$ limit, the result of 't Hooft for the
meson spectrum can also be derived by assuming that the quarks
interact by a linear potential; this condition is equivalent to taking
the Nambu action for a theory of strings connecting the quarks, and
neglecting folds and singular points in the string.  A major tool for
the study of gauge theories was then developed, namely the
Makeenko-Migdal loop equations.  The loop equations for 2-dimensional
gauge theories were first explicitly written down by Kazakov and
Kostov\ftnotemark{3}, who used these relations to compute the
VEV's of Wilson loops on the plane in a large $N$ expansion.  Similar
results were achieved by Bralic using a nonabelian version of Stokes'
theorem\ftnotemark{4}.  The connection between the Wilson loop
VEV's and physical observables of the two-dimensional gauge theory was
made explicit by Strominger, who showed that the Green's functions for
quark bilinears could be explicitly written in terms of an integral
over Wilson loops\ftnotemark{5}.

In a different thread of research, progress was made in understanding
the pure gauge theories in 2 dimensions by exact solution.  It was
first shown by Migdal that by using the heat kernel action for the
gauge theory, one arrives at a theory on the lattice which is
invariant under triangulations, and which is equivalent to the usual
gauge theory as the triangulation becomes arbitrarily
fine\ftnotemark{6}.  Using this approach, it was shown by Rusakov,
Witten, and others that the partition function of the Euclidian gauge
theory on a compact Riemann surface could be exactly expressed in
terms of the group theory of the gauge group\ftnotemark{7}.
Explicitly, they showed that the partition function on a manifold
${\cal M}$ of genus $G$ and area $A$ is given by
\begin{equation}
 Z  (G,\lambda A, N) =   \int[{\cal D} A^\mu]
e^{- {1\over 4 {\tilde g}^2} \int_{\cal M} d^2 x\sqrt{g}\
 Tr F^{\mu \nu}  F_{\mu \nu}}  =  \sum_{R}^{} (\dim R)^{2 - 2G}
  e^{-\frac{\lambda A}{2 N}C_2(R)}, \label{eq:partition}
\end{equation}
where the sum is taken over all irreducible representations of the
gauge group, with $\dim R$ and $C_2(R)$ being the dimension and
quadratic Casimir of the representation $R$. ($\lambda$ is related to
the gauge coupling $\tilde{g}$ by $\lambda = \tilde{g}^2 N$.)
%\vglue 0.58cm
\newpage
{\elevenbf\noindent  3. String Theory}
\vglue 0.38cm
We will now explain how the exact expression Eq.~(\ref{eq:partition}) for
the gauge theory partition function can be rewritten as a string
theory partition function to all orders in $1/N $.
The results in this section were originally described in the
papers [8-10].  Recently, Kostov has described a similar
%	8 9 10
equivalence for a lattice version of the theory\ftnotemark{11}.
\vglue 0.2cm
{\elevenit \noindent 3.1. Statement of Main Result}
\vglue 0.1cm
In the string theory which we describe here, the partition function is
given by a sum over topologically distinct maps from a two-dimensional
string world sheet ${\cal N}$ of any genus $g$ to the target space
${\cal M}$.  Essentially, the string partition function is written
\begin{equation}
Z = \int_{\Sigma({\cal M})} {\rm d} \nu W (\nu),
\label{eq:string}
\end{equation}
where $\Sigma ({\cal M})$ is a set of covering  maps
$\nu:  {\cal N} \rightarrow {\cal M}$,
which are allowed to have singularities at
a finite set of  points in ${\cal M}$.  The
specific types of singular points allowed in the maps in $\Sigma
({\cal M})$ depend upon the choice of gauge group and the genus $G$.
The
weight $W (\nu)$ associated with a certain string map $\nu$
is given by
\begin{equation}
W (\nu) =   \pm
\frac{N^{2-2 g}}{ | S_\nu |}
e^{- \frac{n \lambda A}{2}},
%\label{eq:}
\end{equation}
where $g$ is the genus of ${\cal N}$, $n$ is the degree of the
map $\nu$, and $|S_\nu|$ is the symmetry factor of the map $\nu$ (the
number of diffeomorphisms $\pi$ of ${\cal N}$ which satisfy $\nu \pi =
\nu$).
In general, the manifold ${\cal N}$ need not be connected; the genus
of a disconnected manifold is defined so that the Euler characteristic
is additive for disjoint unions of connected manifolds.  Apart from
the sign, which depends upon the types of singularity points in the
map $\nu$, this is a very natural weight for a string theory.  The
power of $N$ is just the usual power of the string coupling $1/N$, the
exponent of $n \lambda A/2$ is just the usual Nambu action, and the
symmetry factor is the usual one associated with Feynman diagrams in
any field theory.  The unusual feature about this string theory is
that the sum over string maps is restricted to the set of maps $\Sigma
({\cal M})$.  In particular, the strings are not allowed to have folds
other than at the finite number of singular points.  Of course, the
gauge theory we are studying has no propagating degrees of freedom;
allowing folds into the string theory would clearly violate this
characteristic, so the absence of folds is in some sense not
surprising.
\vglue 0.2cm
{\elevenit \noindent 3.2. Outline of Proof -- Simple Case}
\vglue 0.1cm
We will now give a brief description of the essential features in a
proof of Eq.~(\ref{eq:string}); we will also describe in more detail the
set of allowed maps $\Sigma ({\cal M})$.  We will assume for the
remainder of this section that the gauge group is $SU(N)$.  For now,
we will also restrict attention to the torus $G = 1$, where the
contribution from the dimension terms vanishes in
Eq.~(\ref{eq:partition}).  We will return later to the corrections due to
these dimension terms.
%Generally, the structure of the argument
%presented here is that first we rewrite the expression for the
%partition function Eq.~(\ref{eq:partition}) in terms of characters of the
%symmetric group $S_n$.  We can then interpret the resulting equation
%in terms of a topological sum over covering spaces, where the elements
%of the symmetric group describe permutations on sheets of the covering
%space.

In order to write an asymptotic expansion of Eq.~(\ref{eq:partition}) in
powers of $1/N$, we would like to proceed by computing the quadratic
Casimir of each representation $R$ of $SU(N)$ as a function of $N$,
and thus writing an asymptotic expansion separately for each term in
the sum over representations.  Because as $N$ varies, the set of
representations of $SU(N)$ itself changes, we must find a way of
implementing this procedure which is well-defined.  The theory of
Young tableaux gives us such an approach.  We can associate each
representation $R$ with a certain Young tableau, which we also denote
by $R$.  Given a Young tableau $R$ with $n$ boxes, the quadratic
Casimir of the representation is given by
\begin{equation}
C_2 (R)= n N + \frac{n (n - 1)\chi_R (T_n)}{
\chi_R (1)}-\frac{n^2}{N},
\label{eq:casimir}
\end{equation}
where  $\chi_R (T_n)$ and $\chi_R (1)= d_R$ are characters of the
symmetric group $S_n$ in the representation associated with the Young
tableau $R$.  We denote by $T_n$ the conjugacy class of elements in the
symmetric group which have one cycle of length 2 and $n-2$ cycles of
length 1.

We would now like to insert Eq.~(\ref{eq:casimir}) into
Eq.~(\ref{eq:partition}) to form the asymptotic expansion of the
partition function.  However, we must be careful about which
representations we include in the sum.  Naively, one might expect that
in the asymptotic $1/N$ expansion, it would suffice to include all
representations corresponding to Young tableaux with a finite number
of boxes.  In fact, however, it is necessary to include all
representations whose quadratic Casimir is of the form $C_2 (T) = n N
+ {\cal O} (1)$.  In addition to the Young tableaux with a finite
number of boxes, we must include another set of representations
corresponding to the conjugates of these representations, and
``composite'' representations arising {}from tensor products of these
two types of representations.  To simplify the presentation, we will
temporarily assume that the sum in Eq.~(\ref{eq:partition}) can be
replaced by a sum over representations with a finite number of boxes.
We denote this simplified partition function by $Z_Y$.  We will return
to the correct sum over composite representations shortly; the
simplifying assumption of only including Young tableaux with finite
boxes corresponds to only considering a single ``chiral'' sector of
that complete theory.

Replacing the sum over representations by a sum over all Young
tableaux in each set $Y_n$ of tableaux with a finite number $n$ of
boxes, we have
\begin{equation}
Z_Y  (1,\lambda A, N)
=\sum_{n} \sum_{R \in Y_n}
\exp \left[{-\frac{\lambda A}{2 N}(n N + \frac{n (n - 1)\chi_R (T_n)}{
d_R} - \frac{n^2}{N})} \right].
%\label{eq:}
\end{equation}
Expanding the exponential and using some elementary identities from
the theory of characters of the symmetric group,
we have
\begin{eqnarray}
Z_Y (1,\lambda A, N)
 & =  &  \sum_{n, i,t,h}
{\rm e}^{- \frac{n \lambda A}{2}} N^{2-2 g}
\frac{(\lambda A)^{i + t + h}}{i!  \;t!  \; h!}
\frac{(-1)^i n^h (n^{2} - n)^{t} }{2^{t + h}}
\nonumber\\
& &
\cdot
\sum_{p_1, \ldots ,p_i \in T_n}
\sum_{s, t\in S_n} \left[
\frac{1}{n!}   \delta (p_1\cdots p_i
s t s^{-1} t^{-1}
)\right],  \label{eq:expanded}
\end{eqnarray}
where $2-2g =-2 (t + h) - i$.

We would like to now interpret this expression in terms of a sum over
covering maps of the torus.  This interpretation follows from a simple
theorem which holds for any genus $G$.  Define $\Sigma(G,n,i)$ to be
the set of all topologically distinct covering maps onto a genus $G$
target space of degree $n$ and with $i$ elementary branch point
singularities at a fixed set of points $q_j$.  Then
\begin{equation}
\sum_{\nu\in \Sigma(G,n,i)} 1/|S_{\nu}| =
\sum_{p_1, \ldots, p_i \in T_n}
\sum_{s_1, t_1,  \ldots, s_G,t_G \in S_n}
\left[ \frac{1}{n!} \delta (p_1\cdots p_i \prod_{j=1}^{G}  s_j t_j s_j^{-1}
t_j^{-1}) \right].
\label{eq:branched}
\end{equation}
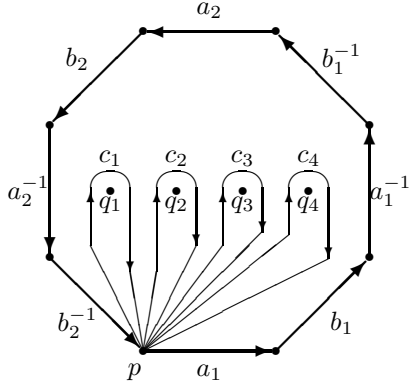
\begin{figure}[htp]
\centering
\begin{picture}(270,270)(-135,-135)
\thicklines
\put(- 50,-121){\vector(1,0){97}}
\put(50,121){\vector(-1,0){97}}
\put(121,- 50){\vector(0,1){97}}
\put(-121,50){\vector(0,-1){97}}
\put( 50,-121){\vector(1,1){68}}
\put(-50,121){\vector(-1,- 1){68}}
\put(121,50){\vector(- 1,1){68}}
\put(-121,-50){\vector(1,-1){68}}
\thinlines
\put(- 50,-121){\circle*{6}}
\put(50,121){\circle*{6}}
\put(121,- 50){\circle*{6}}
\put(-121,50){\circle*{6}}
\put( 50,-121){\circle*{6}}
\put(-50,121){\circle*{6}}
\put(121,50){\circle*{6}}
\put(-121,-50){\circle*{6}}
\put(0,-137){\makebox(0,0){$a_1$}}
\put(100,- 100){\makebox(0,0){$b_1$}}
\put(137,0){\makebox(0,0){$a_1^{-1}$}}
\put(100,100){\makebox(0,0){$b_1^{-1}$}}
\put(0,137){\makebox(0,0){$a_2$}}
\put(-100,100){\makebox(0,0){$b_2$}}
\put(-137,0){\makebox(0,0){$a_2^{-1}$}}
\put(-100,-100){\makebox(0,0){$b_2^{-1}$}}
\multiput(- 75,0)(50,0){4}{\circle*{5}}
\multiput(- 75,0)(50,0){4}{\oval(30,30)[t]}
\put(- 90,- 41){\vector(0,1){41}}
\put(- 50,- 121){\line(-1,2){40}}
\put(- 60,0){\vector(0,-1){61}}
\put(- 50,- 121){\line(-1,6){10}}
\put(- 40,- 61){\vector(0,1){61}}
\put(- 50,- 121){\line(1,6){10}}
\put(- 10,0){\vector(0,-1){41}}
\put(- 50,- 121){\line(1,2){40}}
\put(10,-41){\vector(0,1){41}}
\put(- 50,- 121){\line(3,4){60}}
\put(40, 0){\vector(0,- 1){31}}
\put(- 50,- 121){\line(1,1){90}}
\put(60, -33){\vector(0,1){33}}
\put(- 50,- 121){\line(5,4){110}}
\put(90,0){\vector(0,- 1){51}}
\put(- 50,- 121){\line(2,1){140}}

\put(- 57,- 137){\makebox(0,0){$p$}}
\put(- 75,- 10){\makebox(0,0){$q_1$}}
\put(- 25,- 10){\makebox(0,0){$q_2$}}
\put(25,- 10){\makebox(0,0){$q_3$}}
\put(75,- 10){\makebox(0,0){$q_4$}}
\put(- 75,25){\makebox(0,0){$c_1$}}
\put(- 25,25){\makebox(0,0){$c_2$}}
\put(25,25){\makebox(0,0){$c_3$}}
\put(75,25){\makebox(0,0){$c_4$}}
\end{picture}
\caption[x]{\footnotesize Surface with Genus $G = 2$, $i = 4$ Branch
Points}
\label{f:8gon}
\end{figure}
This theorem can be proven as follows.  We can cut the surface along
the usual homotopy generators $a_j,b_j$ to form a $4G$-gon.  A
covering space  with branch points at $q_j$ can be described by choosing a
labeling $1,2, \ldots, n$ of the sheets of the covering space at a
point $p$, and determining the permutations on this set of labels which
are realized by moving around the homotopy generators and a set of
loops $c_j$ which encircle the branch points.  Denote the
permutations associated with the loops $a_j,b_j$ and $c_j$ by
$s_j,t_j$ and $p_j$ respectively.  The statement that the branch points
are elementary is equivalent to the  condition that $p_j$ is an
element of the conjugacy class $T_n$ for all $j$.  The single condition
on the homotopy group $\pi_1 ({\cal M}\setminus\{q_1, \ldots, q_i\})$
is that
\begin{equation}
c_1 \cdots c_i
a_1 b_1 a_1^{-1} b_1^{-1} a_2 b_2 a_2^{-1} b_2^{-1} \cdots
a_G b_G a_G^{-1} b_G^{-1}=1.
\label{eq:relation}
\end{equation}
The permutation associated with the cycle on the left hand side of
this equation must therefore be the the identity permutation.  We can
now view the sum over all $p_j,s_j,t_j$ in Eq.~(\ref{eq:branched}) as a
sum over all distinct labeled coverings with branch points at $q_j$,
where the $\delta$ function enforces the condition that the
association of permutations to homotopy generators be a homeomorphism
into the symmetric group.  By noting that the number of distinct
labelings for a particular topological type of covering map is
precisely $n!/|S_\nu|$, we see that the combinatorial factor in the sum
over coverings works out correctly, and we have proven
Eq.~(\ref{eq:branched}).

We are now in a position to rewrite the partition function $Z_Y$ in
the form of Eq.~(\ref{eq:string}).  We choose an orientation on ${\cal
M}$, and define a set of covers $\Sigma_+ ({\cal M})$ to be the set of
orientation-preserving covering maps from an oriented Riemann
surface ${\cal N}$ to ${\cal M}$, which have a finite number of branch
point singularities, and in addition a finite number of singularities
corresponding to handles and tubes in ${\cal N}$ which are contracted
to points in ${\cal M}$.  We define a measure ${\rm d} \nu$ on the
space $\Sigma_+ ({\cal M})$ by using the positions of the singular
points as parameters in each connected component, and giving the
position each of singularity a measure proportional to $\lambda {\rm
d}A$.  With respect to this measure, each topological map with $i$
branch points carries a factor of $(\lambda A)^i/i!$, where the
denominator arises from the indistinguishability of the branch points.
Similarly, each map with $h$ contracted handles carries a measure
factor of $(n \lambda A)^h/h!$, and each map with $t$ contracted tubes
carries a measure factor of $(n(n-1) \lambda A)^t/(2^t t!)$.  Each
contracted handle carries a symmetry factor of $1/2$.

Combining these weights, we have the result that the asymptotic
expansion for the partition function on the torus restricted to Young
tableaux with a finite number of boxes can be written
\begin{equation}
Z_Y (1, \lambda A, N) = \int_{\Sigma_+({\cal M})} {\rm d} \nu W (\nu),
\label{eq:stringchiral}
\end{equation}
where the
weight $W (\nu)$
is given by
\begin{equation}
W (\nu) =   (-1)^i
\frac{N^{2-2 g}}{ | S_\nu |}
e^{- \frac{n \lambda A}{2}}.
\label{eq:simpleresult}
\end{equation}
\vglue 0.2cm
{\elevenit \noindent  3.3.  Coupled Theory}
\vglue 0.1cm
As we remarked above, to correctly derive the asymptotic expansion of
the complete theory it is necessary to consider all Young tableaux
with a quadratic Casimir of leading order $N$.  The set of Young
tableaux which satisfy this condition are precisely those tableaux
containing a finite number of columns with $N -k$ boxes where $k$ is
finite, and a finite number of columns with a finite number of boxes.
We will call the representations associated with these tableaux
composite representations.  We write composite representations as
$\bar{S}R$ where $S,R$ are Young tableaux with a finite number of
boxes.  The representation $\bar{S}R$ is constructed by taking the
Young tableau with the maximum number of boxes in the tensor product
of the representation $R$ with the conjugate of the representation
$S$.  Although the quadratic Casimir for any representation which is
not of the composite type scales as $N^2$ and therefore gives a
contribution to the partition function proportional to ${\rm e}^{-N}$,
one might question whether it is really acceptable to neglect such
representations when constructing the asymptotic $1/N$ expansion.
Recent work by Douglas and Kazakov\ftnotemark{12} which will be
discussed in more detail later demonstrates that in fact these
representations can indeed be neglected except on spheres of area
$\lambda A < \pi^2$.

Given Young tableaux $S,R$ with $\tilde{n}, n$ boxes respectively, we
can compute the quadratic Casimir
of the composite representation $\bar{S}R$; we find that
$C_2 ( \bar{S}R) = C_2 (R) + C_2 (S)+ 2n\tilde n/N $.
We can write the complete asymptotic expansion for the partition
function for any genus $G$ as a sum over composite representations,
\begin{equation}
Z  (G,\lambda A, N)  = \sum_{n,\tilde n}
\sum_{R,S \in Y_n, Y_{\tilde n}}^{} (\dim \bar{S}R)^{2 - 2G}
  e^{-\frac{\lambda A}{2 N} \left[  C_2(\bar{S}R)
\right]}.
%\label{eq:}
\end{equation}
%\begin{eqnarray}
%\lefteqn{Z  (G,\lambda A, N)  = }  \nonumber\\
% &  &  \sum_{n,\tilde n}
%\sum_{R,S \in Y_n, Y_{\tilde n}}^{} (\dim \bar{S}R)^{2 - 2G}
%  e^{-\frac{\lambda A}{2 N} \left[  C_2(\bar{S}R)
%\right]}.
%\end{eqnarray}

Except for the term $2n \tilde{n}/N$, this partition function for
genus $G = 1$ factorizes into two components, each equal to the
partition function $Z_Y$.  We can reproduce this factorization
geometrically by including in the set of allowed maps $\Sigma ({\cal
M})$ maps from an oriented Riemann surface ${\cal N}$ which are either
orientation-preserving or orientation-reversing maps, relative to a
fixed orientation on ${\cal M}$.  We refer to these two types of maps
as two ``chiral sectors'' of the string theory.  To see how the
coupling term proportional to $n \tilde{n}$ can be incorporated, one
simply expands the exponential of the quadratic Casimir in the
complete partition function as in Eq.~(\ref{eq:expanded}).  Just as
the terms containing $h$ and $t$ were interpreted as arising from
handles and tubes in the string world-sheet contracted to points in
the target space, we can interpret the contribution from the coupling
term as arising from infinitesimal tubes which connect sheets of
opposite orientation.  The number of ways in which one of these tubes
can connect $n$ sheets of one orientation with $\tilde{n}$ sheets of
the opposite orientation is clearly $n \tilde{n}$, so the counting is
correct.  {}From the sign on the coupling term, we see that each
orientation-reversing tube carries a factor of $-1$.

Thus, in the complete theory containing a sum over all composite
representations, we can define the set of covering maps $\Sigma ({\cal
M})$ to be the set of all maps from a (possibly disconnected) oriented
world-sheet onto ${\cal M}$ which is locally a covering map at all but
a finite number of singular points; the allowed types of singular
points consist of $(i)$ elementary branch points, $(h)$ contracted
handles, $(t)$ contracted tubes between sheets of identical relative
orientation, and $(\tilde{t})$ contracted orientation-reversing tubes.
The partition function is then given by an equation of the form of
Eq.~(\ref{eq:string}),
%\begin{equation}
%Z (1, \lambda A, N) = \int_{\Sigma({\cal M})} {\rm d} \nu W (\nu),
%%\label{eq:string}
%\end{equation}
where the
weight $W (\nu)$
is given by
$W (\nu) =   (-1)^{(i + \tilde{t})}
{N^{2-2 g}}
\exp(- \frac{n \lambda A}{2})/{ | S_\nu |}$.
This completes the  description of the asymptotic $1/N$ expansion of
the $SU(N)$ gauge theory on the torus as a string theory.
\vglue 0.2cm
{\elevenit \noindent  3.4. Example}
\vglue 0.1cm
As a simple example of the equivalence between the gauge theory
partition function and the sum over covering maps, let us consider the
leading order terms in the partition function of a single chiral
sector on the torus.  To order $N^0$, the partition function $Z_Y$ is
given by
\begin{equation}
Z_Y (1,\lambda A, N)= \sum_{n} x^n(\pi_n  + {\cal O}  (1/N))
=\prod_{i}\frac{1}{(1 - x^i)} + {\cal O}(1/N),
%\label{eq:}
\end{equation}
where $x = \exp (- \frac{\lambda A}{2})$, and $\pi_n$ is the number of
ways of partitioning an integer $n$ into a sum of integers (the number
of distinct Young tableaux with $n$ boxes).  In the complete theory
with both chiral sectors, the leading order term is simply given by
squaring this expression.

As in any field theory,  by taking the logarithm of the partition
function, we get a free energy which is given by a sum over connected
diagrams.  In this case,
\begin{equation}
W_Y  (1,\lambda A,N)= \ln Z_Y  (1,\lambda A,N)= \sum_{n} \omega_n x^n
+ {\cal O}(1/N),
%\label{eq:}
\end{equation}
where
$\omega_n =  \sum_{k | n} k/n$.
We expect from the string description of the partition function, that
the quantity $\omega_n$ should be precisely the sum of $1/|S_\nu |$ over
all unbranched connected $n$-fold covers of the torus.  This  equality
can easily be seen to hold, using the fact that all such covers have a
symmetry group of order $n$.  Thus, we have a verification of the
string interpretation in this simple case.
\vglue 0.2cm
{\elevenit \noindent  3.5. Higher Genus}
\vglue 0.1cm
We will now briefly describe the correction to the simple string
theory description which arises when $G \neq 1$ from the insertion of
the $2-2G$th power of the dimension $\dim R$ for each representation.
In terms of the string geometry of the theory, the only change which
occurs from these extra terms can be described by including in the
set $\Sigma ({\cal M})$ maps containing additional singularities at
a set of $|2-2G |$ fixed points in ${\cal M}$.  When $G = 0$ we refer
to these points as $\Omega$-points; when $G > 1$, we call the points
$\Omega^{-1}$-points.  The unusual feature of these points is that
they are not allowed to move on the manifold ${\cal M}$,  and do not
carry factors of the area $A$ as do the other singularities.  In
addition, the types of singularities allowed at $\Omega$-points and
$\Omega^{-1}$-points are different.

In a single chiral sector of the theory, the singularity structure at
an $\Omega$-point is extremely simple.  The map $\nu$ can contain a
singularity which gives rise to an arbitrary permutation on the sheets
of the covering space when one follows a closed loop around the
singularity.  In the coupled theory, at an $\Omega$-point, in each
sector an arbitrary permutation of the sheets of the covering space is
allowed.  However, in addition, the sets of sheets which are connected
by the cycles of these permutations can be connected in pairs by
orientation-reversing tubes.

The set of singularities allowed at a $\Omega^{-1}$-point is very
similar to that of an $\Omega$-point, but slightly more complicated.
Essentially, at an $\Omega^{-1}$-point, there may be an arbitrary
number of nontrivial singularity points of the type from an
$\Omega$-point, each carrying a factor of $-1$.  This description
holds both in the complete theory and in a single chiral sector.
%It is sometimes convenient to think of these separate $\Omega$-point
%type singularities as being separated by an infinitesimal distance on
%${\cal M}$.

In sum, then, by defining the set $\Sigma ({\cal M})$ of covering
spaces to include covers with 2 $\Omega$-point type singularities when
$G = 0$, and to include $2G - 2$ $\Omega^{-1}$-point type
singularities when $G > 1$, we can write the gauge theory partition
function for arbitrary genus as Eq.~(\ref{eq:string}); in this general
case,
%\begin{equation}
%Z (1, \lambda A, N) = \int_{\Sigma({\cal M})} {\rm d} \nu W (\nu),
%%\label{eq:string}
%\end{equation}
$W (\nu)$ is given by
\begin{equation}
W (\nu) =   (-1)^{(i + \tilde{t} + \sum_{j}x_j )}
\frac{N^{2-2 g}}{ | S_\nu |}
e^{- \frac{n \lambda A}{2}},
%\label{eq:}
\end{equation}
where $x_j$ is the number of distinct $\Omega$-point type
singularities at the $j$th $\Omega^{-1}$-point.  Thus, we have defined
a string theory  representation of the partition function for the
$SU(N)$ gauge theory on an arbitrary genus Riemann surface.
\vglue 0.58cm
{\elevenbf\noindent 4. Further Results}
\vglue 0.2cm
{\elevenit \noindent 4.1. Other Gauge Groups}
\vglue 0.1cm
Up to this point, we have restricted attention to the gauge group
$SU(N)$.  However, for other gauge groups it is also possible to
construct a string interpretation of the gauge theory partition
function.  In general, for a given gauge group, it is necessary to
first determine the set of Young tableaux which correspond to
representations whose quadratic Casimir is of order $N$.  By then
expressing the quadratic Casimir and dimension of these
representations in terms of characters of the symmetric group, one can
ascertain the types of singularity structures which are allowed in the
set of string maps $\Sigma ({\cal M})$ associated with that particular
theory.  Generally, the subleading terms in the quadratic Casimir
correspond to ``mobile'' singularity types, which carry factors of the
area, and the subleading terms in the dimension correspond to the
``static'' types of singularities which appear in the $\Omega$-points
of the theory.

The simplest example of this general analysis is for the gauge group
$U(N)$.  For $U(N)$, the expression for the dimension is the same as
for $SU(N)$, so the $\Omega$-points are identical in this theory.  The
quadratic Casimir, however, differs from that for $SU(N)$ in that the
final term $-n^2/N$ associated with vanishing $U(1)$ charge is absent.
Thus, in the $U(N)$ theory the mobile tube and handle singularities do
not occur; on the torus, the only allowed types of singularities are
branch points.

A similar analysis of the string theories for gauge groups $SO(N)$ and
$Sp(N)$ has been carried out by Naculich, Riggs, and Schnitzer, and by
Ramgoolam\ftnotemark{13}.  They found that for these gauge groups, the
string world sheet is not necessarily orientable, and that there are
additional singularities corresponding to infinitesimal cross-caps in
the string maps.  The insertion of a cross-cap singularity essentially
corresponds to cutting out an infinitesimal disk and replacing it with
a projective plane minus a point; moving along a loop which surrounds
this point gives a reversal of orientation.  The singularity structure
at $\Omega$-points for these gauge groups is remarkably similar to
that for $\Omega$-points in the $SU(N)$ theory; however, cross-cap
singularities also appear at these $\Omega$-points.
\vglue 0.2cm
{\elevenit \noindent 4.2. Wilson Loops}
\vglue 0.1cm
Just as the partition function of any 2D gauge theory on an arbitrary
Riemann surface can be written as a weighted sum over closed string
maps, it is possible to show that the VEV of any Wilson loop $\gamma$
can be expressed as a weighted sum over {\em open} string maps, where
the boundary of the string world-sheet is taken to $\gamma$ by the
covering map.  There are some technical complications with the
calculation for Wilson loops with self-intersections; one finds that
each of the disconnected regions into which the Wilson loop cuts the
manifold ${\cal M}$ must be associated with some fixed number of
$\Omega$-point singularities, even for a Wilson loop on the torus.
The details of the calculation of Wilson loop VEV's in terms of open
strings are described in [10].
%In that paper it is also shown
%that evaluation of the VEV's of some simple Wilson loops using the
%string method gives results identical to those achieved by Kazakov and
%Kostov using loop equations in\ftnotemark{14}.
\vglue 0.2cm
{\elevenit \noindent 4.3. Phase Transition}
\vglue 0.1cm
An important related development is the recent demonstration by
Douglas and Kazakov\ftnotemark{12} of a phase transition in the
partition function on the sphere at the point $\lambda A = \pi^2$ (the
trivial small area phase was previously observed by
Rusakov\ftnotemark{14}).  They used techniques familiar from matrix
models to study which representations contribute to the partition
function in the large $N$ regime.  In the phase with $\lambda
A>\pi^2$, they showed that the set of representations which contribute
to the partition function is precisely the set of composite
representations.  Below the critical value of the area,
nonperturbative effects of other representations simplify the
partition function and render invalid the restriction to composite
representations.  This phase transition is analogous to the phase
transition which occurs for the Wilson action at large
$N$\ftnotemark{15}.

The significance of these results for the understanding of QCD through
a string interpretation is not yet clear.  The existence of the phase
transition clearly represents an obstacle to expanding the string
interpretation to small coupling in certain situations; however, the
fact that this phase transition does not occur for higher genus
surfaces or for finite values of $N$ gives hope that it is not a
fundamental obstacle to progress in this direction.
\vglue 0.2cm
{\elevenit \noindent 4.4. Equivalence of QCD${}_2$ to Other Theories}
\vglue 0.1cm
Other recent work has shown that due to its simple group-theoretic
structure, QCD${}_2$ can be related to many other theories of current
interest.  In its Hamiltonian formulation, pure QCD${}_2$ on a
cylinder is essentially equivalent to quantum mechanics on the
manifold of the gauge group.  The theory was recently studied from
this perspective by Douglas\ftnotemark{16}, and by Minahan and
Polychronakos\ftnotemark{17}, and shown to be equivalent to a theory of
free fermions on the circle.  Minahan and Polychronakos showed that by
writing the string formulation of the theory in Hamiltonian form, one
arrives at the bosonization of the fermion theory.  Using the
collective field formalism, they related this theory to a $c = 1$
matrix model.  It has also been shown by several authors that the
introduction of a Wilson loop source in QCD${}_2$ on the cylinder
gives a theory of interacting fermions with a Sutherland-type
interaction\ftnotemark{17,18}.  In a related work, Caselle {\it et
al.} related QCD${}_2$ on the cylinder to a Kazakov-Migdal model with
periodic boundary conditions\ftnotemark{19}.  Their approach was to
explicitly write the heat kernel on the cylinder in terms of the
invariant angles of group elements.  An interesting result of their
analysis is a formula for the grand canonical partition function of
QCD${}_2$ on the cylinder, containing the correct expressions for the
partition function for finite values of $N$.
\vglue 0.2cm
{\elevenit \noindent 4.5. Finite  $N$ Results}
\vglue 0.1cm
Most of the work described so far connecting gauge theories to string
theory has been done in the context of an asymptotic $1/N$ expansion.
For the gauge theories in two dimensions, however, the explicit form
of the exact solution as a sum over representations makes it possible
to consider making such a correspondence for a fixed finite value of
$N$.  For a finite value of $N$, the sum over representations in the
partition function must be restricted to the set of irreducible
representations of the group.  In the case of $SU(N)$, this
corresponds to summing over only Young tableaux with less than $N$
rows.  By extending the group-theoretic analysis described here to the
finite $N$ case, one finds that the effect of this restriction in the
summation can be fairly easily described in the string picture.  The
essential result is that one restricts to only orientation-preserving
string maps, and for any genus one must introduce a new type of static
singularity point.  Only one such point must be introduced for any
genus; at this ``projection'' point, a singularity corresponding to an
arbitrary permutation of the covering sheets can occur.  The weight of
a singularity associated with a permutation $\sigma$ is given by
\begin{equation}
P^{(N)} (\sigma) = \sum_{R \in Y^{(N)}_n} \frac{d_R}{n!} N^{n - K_\sigma}
\chi_R (\sigma),
%\label{eq:}
\end{equation}
where $Y^{(N)}_n$ is the set of Young tableaux with $n$ boxes and less
than $N$ rows, and $K_\sigma$ is the number of cycles in the
permutation $\sigma$.
The details of the construction of the string theory in the finite $N$
$SU(N)$ gauge theory are described in [20].
\vglue 0.58cm
{\elevenbf\noindent  5. Open Questions}
\vglue 0.2cm
{\elevenit \noindent  5.1. Meson Spectrum}
\vglue 0.1cm
It should be possible to reproduce the planar approximation to the
meson spectrum derived by 't Hooft from the string point of view.  The
work of Bars and Hanson\ftnotemark{2}, and of Strominger\ftnotemark{3}
represent a partial result in this direction.  However, using a purely
string-theoretic argument based on the work presented here, one would
like to rederive this result, and calculate the lower order
corrections to this meson spectrum.  This problem presents an
interesting and nontrivial challenge.
\vglue 0.2cm
{\elevenit \noindent  5.2. Action Formulation}
\vglue 0.1cm
The primary reason that we are interested in studying gauge theories
in two dimensions is for the insight which they give into the
structure of gauge theories in 4 dimensions.  Unfortunately, the
formulation presented here of the string theory associated to gauge
theories in two dimensions is highly dependent upon the unique
characteristics of maps from 2-manifolds to other 2-manifolds.  In
order to extend this work to higher dimensions, one natural approach
is to attempt to describe the theory by an action description where
there is no restriction on the string maps allowed.  In such a
formulation, the fact that maps with folds do not contribute to the
partition function would hopefully arise naturally from an integration
over fermion zero modes in the theory or some other familiar
mechanism.  We have several clues to the form that such an action
formulation of the string theory might take.  Clearly, the action will
contain in some way the Nambu action.  The signs associated with
branch points indicate the possible existence of a fermionic structure
in the theory.  The existence of infinitesimal tubes and handles
presents an obstruction to a holomorphic characterization of the
string maps, since no holomorphic map takes an entire handle of a
Riemann surface into a point.  The invariance of the theory under the
group of area-preserving diffeomorphisms indicates that the string
action should also have this invariance.  Finally, the static
singularities associated with $\Omega$-points may give some hint as to
some global structure in the theory we are looking for; the similarity
between the $\Omega$-points for different gauge groups is particularly
striking in this regard.  Unfortunately, as yet no one has
successfully described the string theory presented here with an action
formalism of the desired type, or even made clear progress towards
such a formalism.
\newpage
{\elevenbf\noindent  7. References}
\baselineskip=13pt
\vglue 0.38cm

\end{document}